\documentclass[pdftex]{pasj01}
\Received{$\langle$reception date$\rangle$}
\Accepted{$\langle$acception date$\rangle$}
\Published{$\langle$publication date$\rangle$}
\newcommand{\object}{CXOU~J171405.7--381031}
\usepackage{color} 
\usepackage{graphicx} 

\begin{document}

\title{XMM-Newton Spectrum of the magnetar CXOU~J171405.7--381031}
\author{Haruka Watanabe,$^{1}$ Aya Bamba,$^{2,3}$ Shinpei Shibata,$^{4}$ and Eri Watanabe$^{1}$}%
\altaffiltext{1}{School of Science and Technology, Yamagata University, 1-4-12 Kojirakawa, Yamagata  990-8560, Japan}
\altaffiltext{2}{Department of Physics, The University of Tokyo, 7-3-1 Hongo, Bunkyo-ku, Tokyo 113-0033, Japan}
\altaffiltext{3}{Research Center for the Early Universe, School of Science, The University of Tokyo, 7-3-1 Hongo, Bunkyo-ku, Tokyo 113-0033, Japan}
\altaffiltext{4}{Department of Physics, Yamagata University, 1-4-12 Kojirakawa, Yamagata  990-8560, Japan}
\email{shibata@sci.kj.yamagata-u.ac.jp}


\maketitle

\begin{abstract}
We observe the magnetar \object\ with XMM-Newton and obtain
the most reliable X-ray spectral parameters for this magnetar.
After removing the flux from the surrounding supernova remnant CTB~37B,
the radiation of \object\ is best described by a two-component model,
consisting of a blackbody and power law.
We obtain a blackbody temperature of 
$0.58^{+0.03}_{-0.03}$~keV,
photon index of
$2.15^{+0.62}_{-0.68}$,
and unabsorbed 2--10 keV band flux of
$2.33^{+0.02}_{-0.02} \times 10^{-12}$ erg~cm$^{-2}$~s$^{-1}$.
These new parameters enable us to compare \object\ with other magnetars, and 
it is found that
the luminosity, temperature and the photon index of \object\ 
are aligned with the known trend among the magnetar population
with a slightly higher temperature, which could be caused by its young age.
All the magnetars with a spin-down age of less than 1~kyr show
time variation or bursts except for \object .
We explore the time variability for 
ten observations in between 2006 and 2015,
but there is no variation larger than $\sim 10$\%.
\end{abstract}

\onecolumn
\section{Introduction}

Magnetars are strongly magnetized neutron stars.
Their magnetic fields, 
which are inferred from the rotation period and its time derivative,
typically exceed the quantum critical field strength
$B_{\rm Q}=m^2 c^3/\hbar e =4.4 \times 10^{13}$~G.
Most of their luminosity is in the X-ray bands, and
the origin of the radiation is believed to be the decay of the strong magnetic field,
although its mechanism and radiation process are not well understood
(for reviews, see  Turolla et al.  2015, Kaspi \& Beloborodov  2017).
X-ray spectra of magnetars in quiescence are generally multi-component;
therefore, 
it is important to obtain precise spectra to distinguish the 
components and clarify the emission mechanism.
The standard spectral model for magnetars is
a blackbody plus power law model (BB$+$PL) in the 0.5--10~keV bands,
although two blackbody models (2BB) are sometimes used.
The Comptonized blackbody model (CBB) or an appropriate neutron-star 
atmosphere model are also sometimes useful.
The spectrum turns upward in the higher energy bands
$\gtrsim 10$~keV (Kuiper et al. 2004), so
X-ray radiation from magnetars is generally decomposed 
into a soft thermal component and a hard non-thermal component.
Enoto et al.(2010) reported that
the ratio of the soft thermal flux and the hard non-thermal flux,
$F(15-60~\mbox{keV})/F(1-10~\mbox{keV})$,
 is negatively correlated with the spin-down age 
and positively correlated with the dipole field strength. 
More importantly, the magnetars show time variability on various time scales 
from milliseconds to seconds (outbursts) and also on longer time scales
(e.g., Kaspi \& Beloborodov  2017).
In this paper, we report detailed spectroscopy of the magnetar 
\object\ with XMM-Newton and
time variability over a decade.

\object\  was found as an X-ray point source associated 
with the shell-type supernova remnant (SNR) CTB~37B
(Aharonian et al. 2008)
in the Chandra follow-up observation for the TeV source
HESS~J1713--38 (Aharonian et al. 2006).
The SNR is also detected in the GeV band with Fermi LAT
(Xin et al. 2016).
Sato et al. (2010) and Halpern \& Gotthelf (2010b) found
the pulse period and its time derivative, and they concluded that
\object\ is a magnetar.
The period and its time derivative are $P=3.825$ s and
$\dot{P}=6.4 \times 10^{-11} {\rm s}\; {\rm s}^{-1}$, respectively,
which provide the spin-down luminosity of 
$L_{\rm rot} =4.5 \times 10^{34}$erg s$^{-1}$,
a dipole field of $B_d = 5.0 \times 10^{14}$~G and 
a characteristic age of $\tau = 0.95$~kyr.
These parameters make \object\ 
the fourth youngest magnetar known and 
the fourth strongest magnetic field known. 
The distance is estimated to be $\sim 13.2$~kpc (Tian \& Leahy 2012).

The spectrum of \object\ was not well determined
by the previous Chandra observation made
on 25 January 2009 and 30 January 2010
(Halpern \& Gotthelf  2010a, 2010b).
Sato et al.(2010) observed \object\ with XMM-Newton on
17--18 March 2010 and analysed only the pn data.
They fitted the data by a power law model only.
In this paper, we use MOS data and pn data
and apply multi-component
spectral models that are standard for magnetars.

\section{Analysis}

\subsection{Observation and data reduction}

\object\ was observed with XMM-Newton
from 13:16:16, 17 2010 March 
to 23:06:33, 18 March 2010 (ObsID = 0606020101).
For the spectral analysis we used data
from the European Photon Imaging Camera (EPIC) instruments:
MOS (Tumer et al. 2001) and pn (Str\"{u}der et al. 2001).
Both the MOS and the pn instruments were configured in full-frame mode,
with a medium filter.

Data reduction was performed with SAS version 16.1.0 and the SAS guide.
The data reduction was done with SAS version 16.1.0 and the SAS guide.
We filtered out background flare times for rates 
in the 10--12~keV band higher than 0.35~cnts~s$^{-1}$,
resulting in 
a total exposure of 100.4~ks for MOS1, 100.7~ks 
for MOS2 and 46.8~ks for pn.
For the following analysis, we used heasoft 6.18 package.
We also used Chandra data and short-exposure XMM-Newton data for the flux monitoring.
Chandra observed \object\ four times 
with ACIS-S in the continuous clocking mode and 
once with ACIS-I in Timed Exposure modes in very faint telemetery format.
The data reduction was done with CIAO version 4.9,
and the total exposure was $\sim 20$~ks for each data set (Table~\ref{tablog}).
After 2010, XMM-Newton observed three times in the small window mode with
the exposure time of $\sim 20$~ks, and these data are used for the flux
monitoring.
The details of the observations are 
summarized in Table~\ref{tablog}.

\begin{table}
\tbl{Observation log of \object\ \label{tablog}}{%
\begin{tabular}{lccc} \hline\hline
Date (YYYY/MM/DD) & Satellite & ObsID & Exposure (ks) \\ \hline
2010/03/17 & XMM-Newton & 0606020101 & 100 (MOS) \\
& & & 45 (pn) \\
2007/02/02 & Chandra & 6692  & 25 (ACIS-I) \\
2009/01/25 & Chandra & 10113 & 30 (ACIS-S) \\
2010/01/30 & Chandra & 11233 & 26 (ACIS-S) \\
2012/07/16 & Chandra & 13749 & 20 (ACIS-S) \\
2015/10/13 & Chandra & 16763 & 20 (ACIS-S) \\
2012/03/13 & XMM-Newton & 0670330101 & 13 (MOS) \\
2016/09/23 & XMM-Newton & 0790870201 & 29 (MOS) \\
2017/02/22 & XMM-Newton & 0790870301 & 22 (MOS) \\
\hline
\end{tabular}}
\end{table}

\subsection{Spectral analysis}

\begin{figure}
        \begin{center}
\includegraphics[width=120mm]{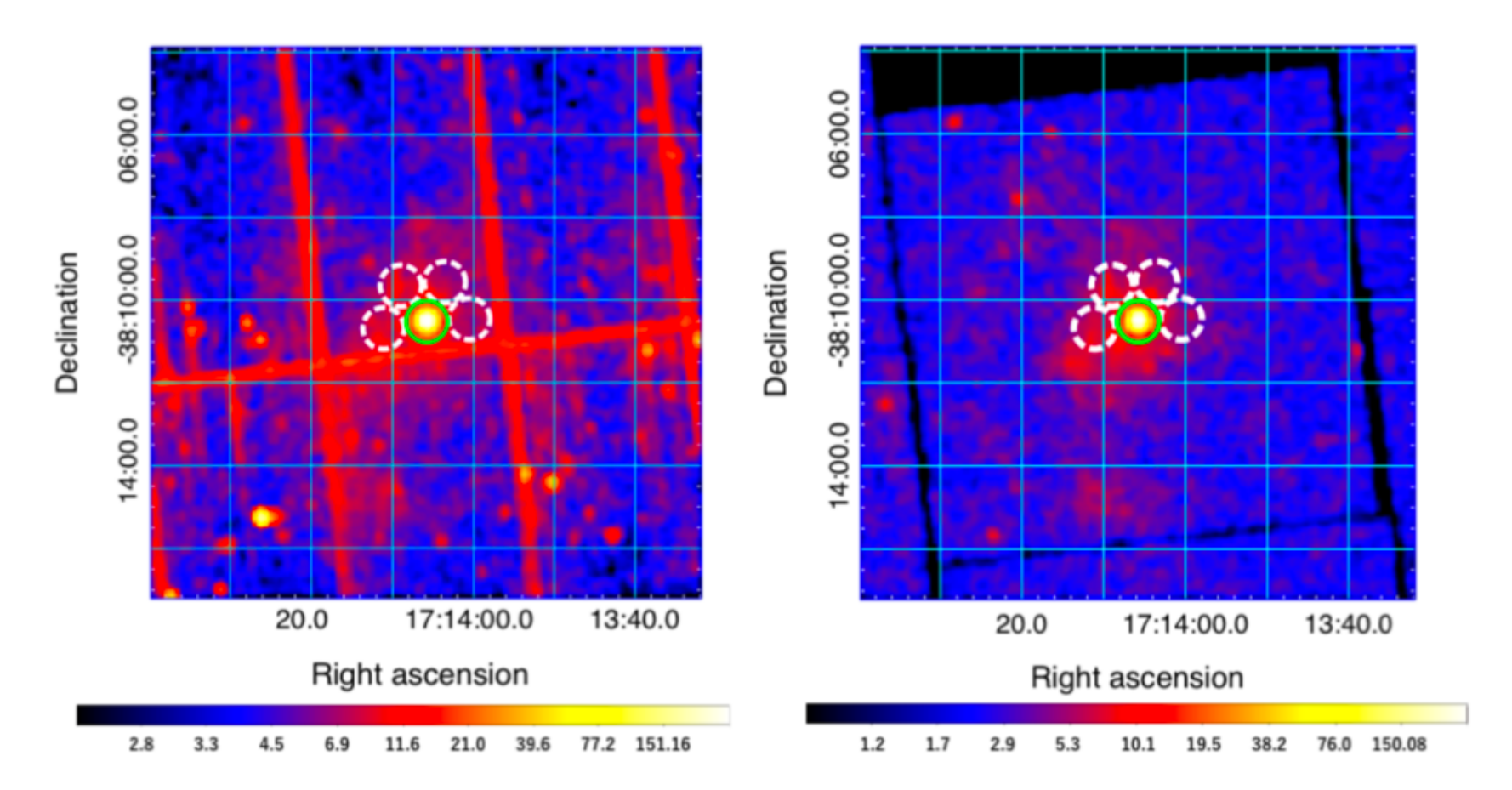}
\caption{\label{image}
XMM-Newton MOS (right panel) and pn (left panel) images
of the CTB~37B region in the 1-10~keV band. 
The source regions for \object\ are indicated by the green circles, 
and the background regions are indicated by the dashed white circles.}
        \end{center}
\end{figure}

Figure~\ref{image} shows the MOS and pn images of the SNR CTB~37B region.
\object\ is detected as a bright point source near the center of 
CTB~37B.
For our spectral analysis, we took the source region as
a circular region
centered at $\ell = 348^\circ 40^\prime 51^{\prime \prime}$,
$b = 0^\circ 22^\prime 15^{\prime \prime}.820$ with a radius of
$30^{\prime \prime}$ (Sato et al. 2010).

\object\ is located within the SNR, so 
the SNR counts must be subtracted from the source counts.
We took background regions near the source, 
namely
the four circular regions with the same radius as the source region,
indicated by the dashed circles in Figure~\ref{image}.
For MOS and pn,
the background regions were set to avoid
the edges of the CCD chips.
We used these four regions together
to obtain the background-subtracted source counts.
The background contribution is 
$\sim 0.08$\% of the source counts in the 1-10~keV band.

An acceptable spectral fit was not obtained with a single blackbody 
or a single power law.
Two-component models, 
either the 2BB model 
or the BB$+$PL model, 
generally give a good fit for magnetars.
The 2BB model
with interstellar absorption (phabs: Balucinska-Church \& McCammon 1992)
gave a reasonably good fit to the present data set with $\chi^2$/d.o.f = $332.39/306$.
The hydrogen column density,
lower and higher blackbody temperatures, and unabsorbed flux are
$n_H = 2.50^{+0.11}_{-0.10} \times 10^{22}\; {\rm cm}^{-2}$,
$kT_1=0.61^{+0.02}_{-0.02}$~keV,
$kT_2=2.10^{+0.50}_{-0.33}$~keV,
and
$F_{\rm x}=2.21^{+0.02}_{-0.02}\times 10^{-12}$~erg~cm$^{-2}$~s$^{-1}$
in the 2--10 keV band,
respectively.
The spectral parameters are summarized in Table~\ref{tabspec}.
The errors are given at the 90\% confidence level, 
and this is the same throughout
this paper.

\begin{figure*}
        \begin{center}
\includegraphics[width=120mm]{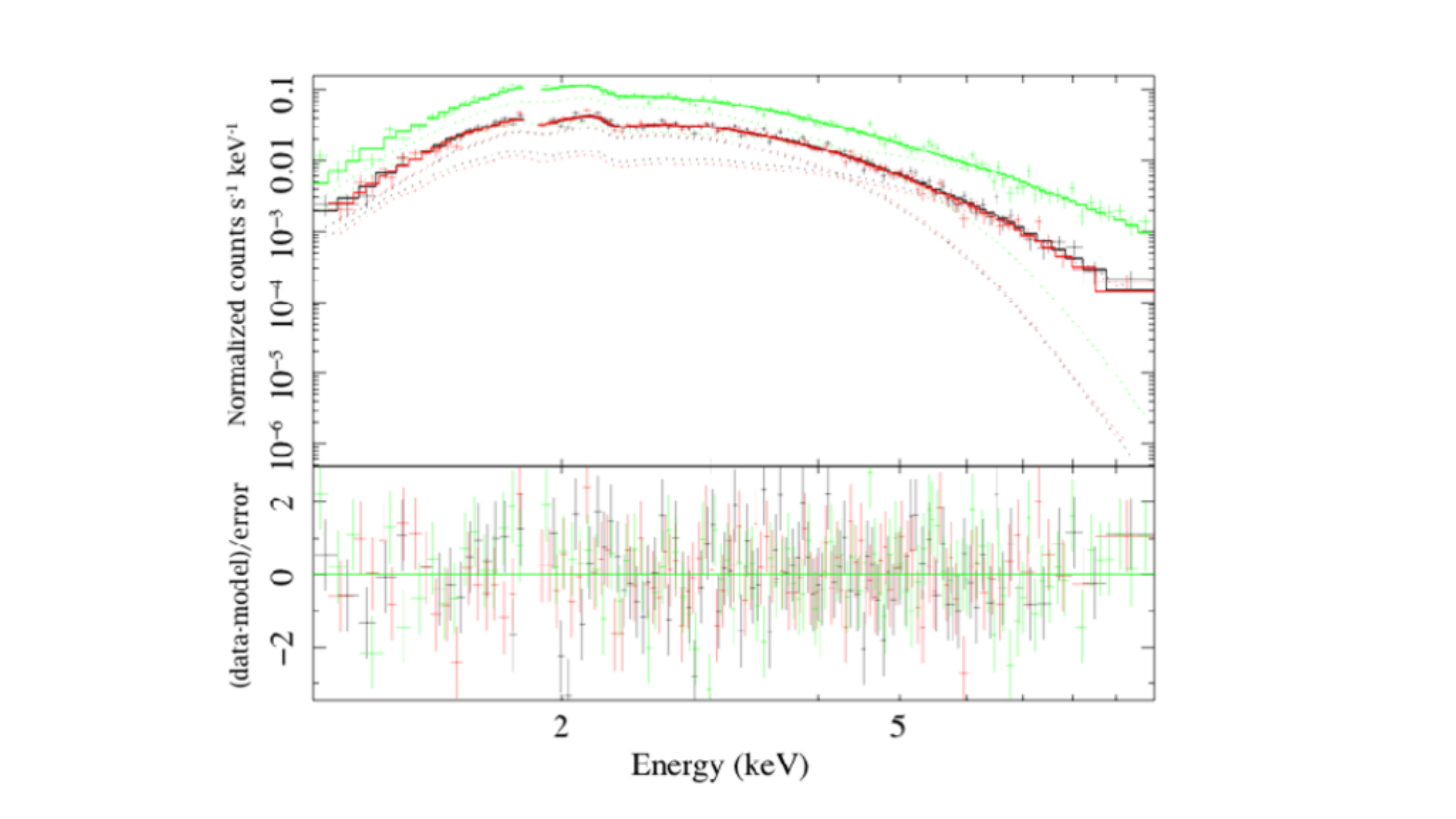}
\caption{\label{spec} XMM-Newton MOS and pn spectra of \object\
The top panel shows the data (crosses) and best fit model (solid line) 
for the parameter given in Table~\ref{tabspec}. 
The spectral components are shown by the dashed lines.
The data for MOS1, MOS2 and pn are shown in black, red and green, 
respectively.
The bottom panel shows the residuals.
}
        \end{center}
\end{figure*}

We also obtained a reasonably good fit by using
the absorbed BB$+$PL model, 
where the same absorption model (phabs) is used.
The chi squared value was $\chi^2$/d.o.f = $325.40/306$, 
which was slightly better than that of the 2BB fit.
Figure~\ref{spec} shows the background-subtracted spectrum of \object\ 
together with the fitted model spectrum and the residuals.
MOS1, MOS2 and pn data were jointly fitted by a single model.
The hydrogen column density is 
$n_H = 2.84^{+0.43}_{-0.28} \times 10^{22}\; {\rm cm}^{-2}$,
and
the blackbody temperature, photon index and unabsorbed flux are
$kT=0.58^{+0.03}_{-0.03}$~keV,
$\Gamma = 2.15^{+0.62}_{-0.68}$,
$F_{\rm x}=2.33^{+0.02}_{-0.02}\times 10^{-12}$~erg~cm$^{-2}$~s$^{-1}$
in the 2--10 keV band,
respectively.
The results are summarized in Table~\ref{tabspec}.

We evaluated systematic uncertainties 
due to inhomogeneous background emission from the SNR. 
We subtracted each of the four background counts from the source counts
to examine how the spectral parameters vary.
We find that the systematic errors 
for the obtained $n_H$, $\Gamma$ and $kT$
for the BB$+$PL model
are
$0.45 \times 10^{22}\; {\rm cm}^{-2}$,
$0.64$,
$0.037$~keV,
at the 90\% confidence level,
which are 130\%, 98\% and 130\% of the statistical error, 
respectively. 
Thus the systematic error due to the inhomogeneity 
of the SNR was comparable to
the statistical error.

\begin{table}
\tbl{Spectral fits to \object\ \label{tabspec} }{%
\begin{tabular}{llll} \hline \hline
Parameter & 2BB & BB$+$PL & CBB \\ \hline
chi-Squared/d.o.f & 332.39/306 & 325.40/306 & 329.67/309 \\
Hydrogen column density$^{*1}$ & $2.50_{-0.10}^{+0.11}$ & $ 2.84_{-0.28}^{+0.43}$ & $2.63^{+0.10}_{-0.10}$ \\
$kT_1$(keV) & $0.61_{-0.02}^{+0.02}$ & $0.58_{-0.03}^{+0.03}$ & $0.56^{+0.02}_{-0.02}$ \\
$R_1$(km)   & $1.82_{-0.01}^{+0.01}$ & $1.83_{-0.02}^{+0.01}$ & ... \\
$kT_2$(keV) & $2.10_{-0.33}^{+0.50}$ & ... &  ...\\
$R_2$(km)   & $0.084_{-0.001}^{+0.002}$ & ... & ... \\
Photon index & ... & $2.15_{-0.68}^{+0.62}$ & ... \\
Optical depth & ... & ... & $0.43^{+0.02}_{-0.02}$ \\
Observed flux$^{*2}$ & $1.58^{+0.02}_{-0.02}$ & $ 1.58^{+0.02}_{-0.02}$ & $1.58^{+0.02}_{-0.02}$ \\
Intrinsic flux$^{*2}$ & $2.21^{+0.02}_{-0.02}$ & $2.33^{+0.02}_{-0.02}$ & $2.25^{+0.02}_{-0.02}$ \\
\hline
\end{tabular}}
\begin{tabnote}
*1 in units of $10^{22}$cm$^{-2}$

*2 in units of $10^{-12}$ erg  cm$^{-2}$ s$^{-1}$
in the 2--10~keV band
\end{tabnote}
\end{table}

In the magnetosphere of persistent magnetars, 
high-energy particles are thought to be supplied continuously 
into magnetic loops, so 
the hard component can be interpreted as a Comptonized tail of 
the blackbody radiation by these high-energy particles  
(Turolla et al., 2015).
We also attempted the fitting with the CCB model.
The results are summarised in Table~\ref{tabspec}.
The electron temperature was assumed to be 50~keV and 100~keV, 
and both gave a good fit with similar blackbody temperatures.
In this case, an additional power component was not needed.
Notably, the standard two-component model and CBB model
give similar temperatures.
The best fit was given by BB$+$PL, followed by CB and 2BB.

We explored the long-term flux variation for \object\ because
this class of magnetars, 
which has a characteristic age smaller than 1~kyr
shows burst activity or variability.

\begin{figure}
        \begin{center}
\includegraphics[width=75mm]{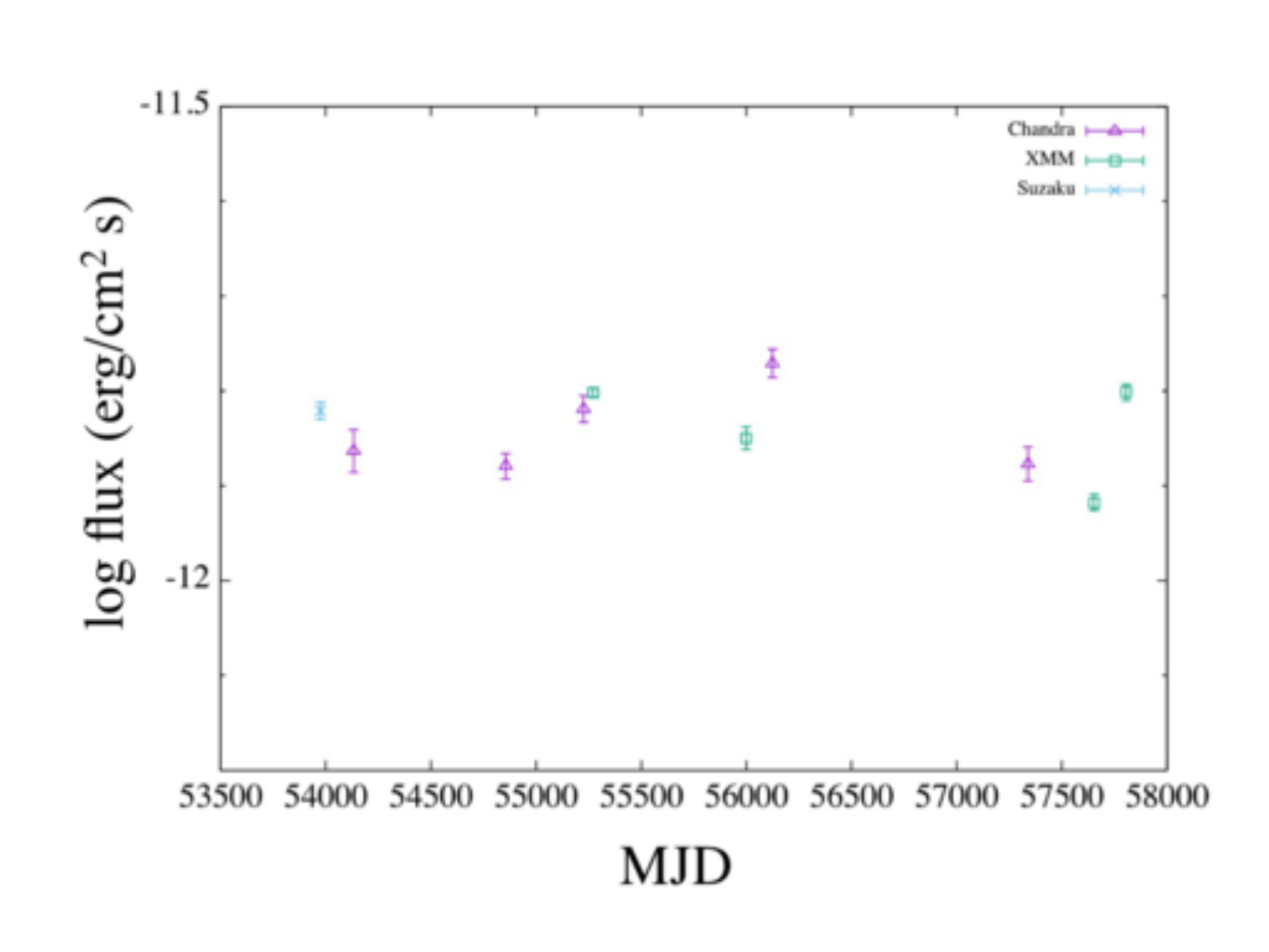}
\caption{\label{lightc} The long term history of the absorbed flux of \object\ in
the $2-10$~keV bands. }
        \end{center}
\end{figure}

We reanalyzed the previous Chandra observations 
and XMM-Newton observations
using the present BB$+$PL spectral parameters, 
which were fixed except for the normalization,
to determine the flux variability.
The $\chi^2$ values were stable and acceptable for all the data, 
and the flux variability results 
are summarized in Table~\ref{tabvar} and Figure~\ref{lightc}.
We found no significant variation in the flux for about a decade.
The mean value of the flux in the 2--10 keV band  was
$1.45 \times 10^{-12}$~erg~cm$^{-2}$~s$^{-1}$ with
the standard deviation of
$0.14 \times 10^{-12}$~erg~cm$^{-2}$~s$^{-1}$ (9.7\% of the mean).

\begin{table}
\begin{center}
\tbl{Long term variation of \object\   \label{tabvar}}{%
\begin{tabular}{lrccl} \hline
Date &flux$*1$(2-10keV)&Model& Mission & Refs \\ \hline
\hline 
 2006/08/27-29 & $ 1.510^{+ 0.030}_{- 0.030}$ & PL & Suzaku & 1 \\ 
 2007/02/02    & $ 1.371^{+ 0.072}_{- 0.072}$ & BB$+$PL & Chandra & 2 \\ 
 2009/01/25    & $ 1.320^{+ 0.041}_{- 0.041}$ & BB$+$PL & Chandra & 2 \\ 
 2010/01/30    & $ 1.517^{+ 0.049}_{- 0.049}$ & BB$+$PL & Chandra & 2 \\ 
 2010/03/17-18 & $ 1.578^{+ 0.017}_{- 0.017}$ & BB$+$PL & XMM & 2 \\ 
 2012/01/16    & $ 1.695^{+ 0.060}_{- 0.059}$ & BB$+$PL & Chandra & 2 \\ 
 2012/03/13    & $ 1.414^{+ 0.040}_{- 0.039}$ & BB$+$PL & XMM & 2 \\ 
 2015/11/13    & $ 1.329^{+ 0.055}_{- 0.055}$ & BB$+$PL & Chandra & 2 \\ 
 2016/09/23    & $ 1.209^{+ 0.023}_{- 0.023}$ & BB$+$PL & XMM & 2 \\ 
 2017/02/22    & $ 1.581^{+ 0.031}_{- 0.031}$ & BB$+$PL & XMM & 2 \\ 
\hline 
\end{tabular}}
\begin{tabnote}
*1 absorbed flux in units of $10^{-12}$~erg~cm$^{-2}$~s${-1}$ \\
References,
1:  Nakamura et al.(2009),
2: this work.
\end{tabnote}
\end{center}
\end{table}

\section{Discussion}

In this paper, 
we have obtained the most reliable spectral parameters for 
\object\ so far by using XMM-Newton.
In the previous Chandra ACIS-I/TE mode observation 
on 25 January 2009 
the data showed a rising spectrum above $>6$~keV, such that
the 2BB fit gave an unphysical high temperature 
and the BB$+$PL fit gave a photon index of $-2$, which is unusual 
(Halpern \& Gotthelf, 2010a).
These high-energy events disappear in the Chandra CC mode observation
on 30 January 2010 (Halpern \& Gotthelf 2010b).
They argued that the pulsar was dithered into the gap between CCD's. 
Thus, the spectral fit for \object\ was uncertain.
We reanalyzed the Chandra data in the previous
section and find that our 2BB parameters also give a good fit
with $\chi^2$/dof $=131.86/135$. 
The Chandra result (Halpern \& Gotthelf, 2010b)
 gave a $\sim 1.4$ times higher $T_1$ and 
a $\sim 1.5$ times larger hydrogen column density than 
the present XMM result.
The difference is caused by how the high energy part is fit, and
the Chandra data does not give enough statistics in the high energy part.
The XMM-Newton observation on 17--18 March 2009 was analyzed 
using only the pn data with a single PL model (Sato et al., 2010). 
In this paper, we analyze both MOS and pn data to give
the spectral parameters for the two-component models that are 
commonly used for magnetars.

In the case of the 2BB fit,
the ratio of the two temperatures of each components shows 
a correlation
over a large number of both persistent and transient magnetars,
as seen in Fig.~3 of Nakagawa et al. (2009), namely,
$kT_2 / kT_1 = 2.7 \pm 1.1$.
The present value of
$3.4$ ($=2.10$~keV$/0.61$~keV) 
follows this correlation.
They suggested that the ratio of the emitting area is
$R_2^2 / R_1^2 \approx 0.01$, whereas 
we find $R_2^2 / R_1^2 = 0.02$, and 
this is again along the general trend in the scatter
(see also Fig.~3 of Nakagawa et al., 2009).

\begin{figure}
        \begin{center}
\includegraphics[width=75mm]{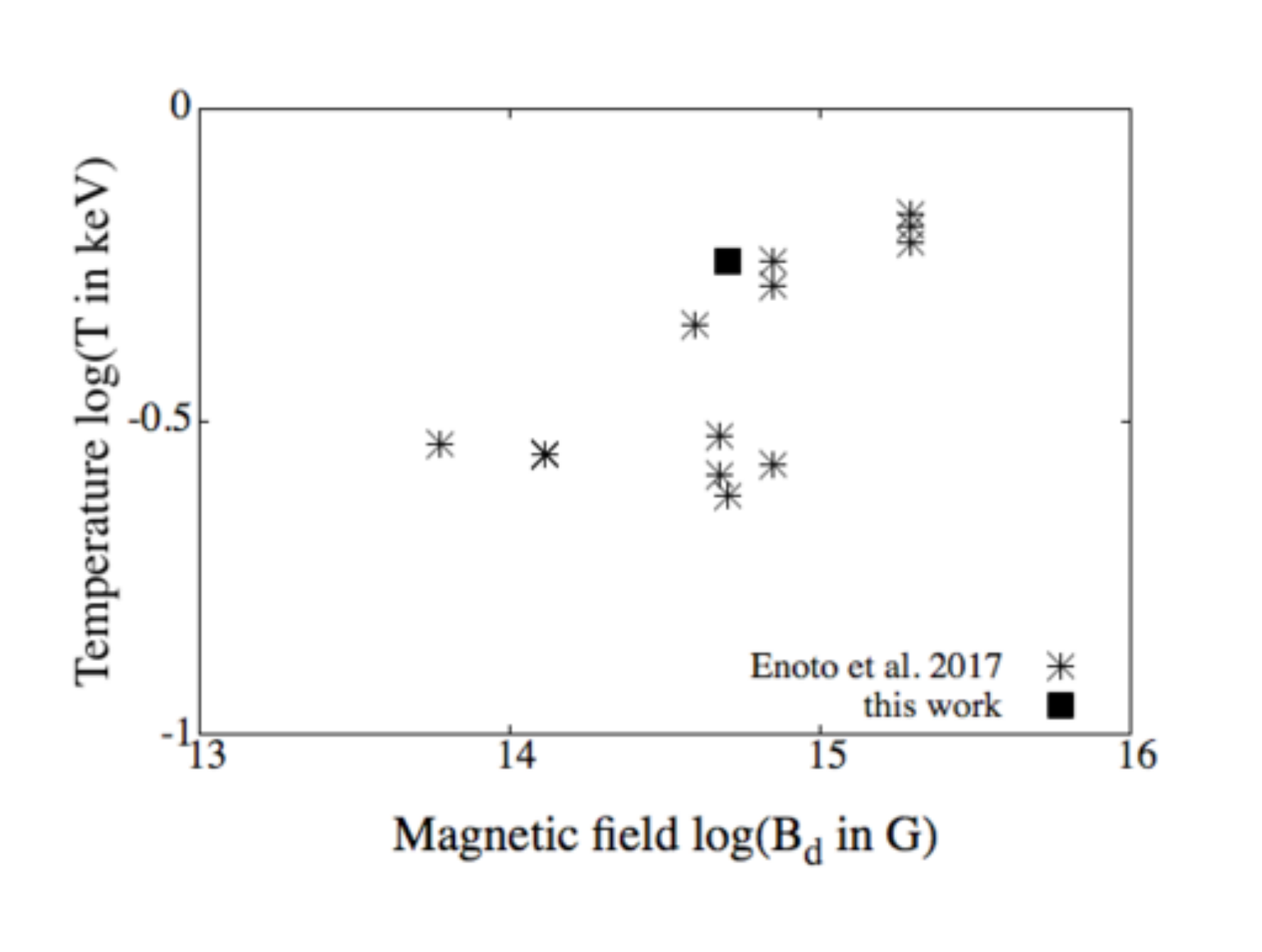}
\caption{\label{ktB} $kT$ versus $B_d$ for the persistent emission of the magnetars
(Enoto 2017) and the present result for \object . }
        \end{center}
\end{figure}

As shown in the previous section, the BB$+$PL model gives the best
fit giving
the X-ray luminosity in the $2-10$~keV band as 
$L_{\rm x}=4.9 \times 10^{34} d^2_{13.2}$~erg~s$^{-1}$, which 
exceeds the spin-down power and is a typical
value for persistent magnetars.
The blackbody component gives an intrinsic bolometric flux of 
$2.34 \times 10^{-12}$~erg~cm$^{-2}$~s$^{-1}$ and 
an emitting radius of $1.8 \; d_{13.2}$~km at
a distance of 13.2~kpc, to which the normalized distance,
$d_{\rm 13.2}$, refers.
For the persistent magnetars,
a positive correlation between the blackbody temperature and 
the dipole field has been suggested
(Pons et al.,2007; Enoto et al. 2017).
Although it was also suggested that 
the correlation is not solid due to a large scatter (Zhu et al., 2011),
the scatter seems natural because the surface magnetic field
has multi-polar components. 
In other words, the magnetic field is 
composed of  many magnetic loops,
and therefore the dipole field
does not always represent the surface magnetic field 
but would still be related to the surface field.
Figure~\ref{ktB} shows the scatter plot of the blackbody temperature and
the dipole field strength obtained by
Enoto et al. (2017), showing a positive correlation. 
Our result for \object\ follows this trend, but it is at
a higher position than the general trend. 
This may be because \object\ is the fourth youngest magnetar known.

The hard power law component of the magnetar spectrum
extends to the soft gamma-ray bands with photon a index 
between 0 to 2.
The photon index shows a positive correlation with the magnetic field 
(Enoto et al., 2010: Kaspi \& Boydstun, 2010).
We obtained a photon index of 1.9 for {\object}, 
which follows this correlation.
However, 
the hard power component becomes dominant beyond $\sim 10$~keV,
which is outside the present observations.
Therefore, the photon index obtained in this works 
is not that of the hard power law component.
In fact, the CBB model does not require the power law component.
Future observations above 10~keV are required.

The magnetar population consists of
persistent and transient sources.
The typical luminosity of the persistent sources is
$10^{33}-10^{35} {\rm erg} \ {\rm s}^{-1}$.
The transient sources brighten to this range of luminosity 
in burst phases,
but generally they dim to levels that cannot be
observed with the current instruments.
The origin of the difference between persistent and transient sources
is unknown.
In this work, 
we did not find any significant change in luminosity
for {\object}.
However, the four magnetars that have characteristic 
ages of less than 1~kyr,
SGR~1806--20, 1E~1841--045, SGR~1900$+$14, and 1E~1547.0--5408, all
show outbursts (Coti Zelati et al., 2018).
Consequently, we still need to monitor this object.

\begin{ack}
We acknowledge support from JSPS/MEXT KAKENHI grant numbers
18H01246 (SS), 15K051017 (AB) and JP18H05459 (AB).
This work was also supported by the Grant-in-Aid for Scientific Research on
Innovative Areas ``Toward new frontiers : Encounter and synergy of
state-of-the-art astronomical detectors and exotic quantum beams'' (AB).
\end{ack}


\begin{thebibliography}{}
\bibitem[Aharonian et al.(2008)]{2008A&A...486..829A} Aharonian, F., Akhperjanian, A.~G., Barres de Almeida, U., et al.\ 2008, \aap, 486, 829
\bibitem[Balucinska-Church \& McCammon(1992)]{1992ApJ...400..699B} Balucinska-Church, M., \& McCammon, D.\ 1992, \apj, 400, 699 
\bibitem[Coti Zelati et al.(2018)]{2018MNRAS.474..961C} Coti Zelati, F., Rea, N., Pons, J.~A., Campana, S., \& Esposito, P.\ 2018, \mnras, 474, 961 
\bibitem[Enoto et al.(2010)]{2010ApJ...722L.162E} Enoto, T., Nakazawa, K., Makishima, K., Rea, N., Hurley, K., \& Shibata, S.\ 2010, \apjl, 722, L162
\bibitem[Enoto et al.(2017)]{2017ApJS..231....8E} Enoto, T., Shibata, S., Kitaguchi, T., et al.\ 2017, \apjs, 231, 8
\bibitem[Halpern \& Gotthelf(2010a)]{2010ApJ...710..941H} Halpern, J.~P., \& Gotthelf, E.~V.\ 2010a, \apj, 710, 941
\bibitem[Halpern \& Gotthelf(2010b)]{2010ApJ...725.1384H} Halpern, J.~P., \& Gotthelf, E.~V.\ 2010b, \apj, 725, 1384 
\bibitem[Kaspi \& Beloborodov(2017)]{2017ARA&A..55..261K} Kaspi, V.~M., \& Beloborodov, A.~M.\ 2017, \araa, 55, 261
\bibitem[Kaspi \& Boydstun(2010)]{2010ApJ...710L.115K} Kaspi, V.~M., \& Boydstun, K.\ 2010, \apjl, 710, L115 
\bibitem[Kuiper et al.(2004)]{2004ApJ...613.1173K} Kuiper, L., Hermsen, W., \& Mendez, M.\ 2004, \apj, 613, 1173
\bibitem[Nakagawa et al.(2009)]{2009PASJ...61..109N} Nakagawa, Y.~E., Yoshida, A., Yamaoka, K., \& Shibazaki, N.\ 2009, \pasj, 61, 109。
\bibitem[Nakamura et al.(2009)]{2009PASJ...61S.197N} Nakamura, R., Bamba, A., Ishida, M., et al.\ 2009, \pasj, 61, S197
\bibitem[Pons et al.(2007)]{2007PhRvL..98g1101P} Pons, J.~A., Link, B., Miralles, J.~A., \& Geppert, U.\ 2007, Physical Review Letters, 98, 071101
\bibitem[Sato et al.(2010)]{2010PASJ...62L..33S} Sato, T., Bamba, A., Nakamura, R., \& Ishida, M.\ 2010, \pasj, 62, L33
\bibitem[Turolla et al.(2015)]{2015RPPh...78k6901T} Turolla, R., Zane, S., \& Watts, A.~L.\ 2015, Reports on Progress in Physics, 78, 116901 
\bibitem[Xin et al.(2016)]{2016ApJ...817...64X} Xin, Y.-L., Liang, Y.-F., Li, X., et al.\ 2016, \apj, 817, 64
\bibitem[Zhu et al.(2011)]{2011ApJ...734...44Z} Zhu, W.~W., Kaspi, V.~M., McLaughlin, M.~A., et al.\ 2011, \apj, 734, 44
\end{thebibliography}
\end{document}